\documentclass[cits]{PoS}

\usepackage{amsmath}

\title{All-order colour structure in threshold resummation and
  squark-antisquark  production at NLL}

\ShortTitle{Colour structure in threshold resummation and
  squark-antisquark  production}
\author{Martin Beneke \\
Institut f\"ur Theoretische Physik E, 
RWTH Aachen University,\\  D - 52056 Aachen, Germany
}
\author{Pietro Falgari\\
IPPP, Department of Physics, University of Durham, \\
Durham DH1 3LE, England
}
\author{\speaker{Christian Schwinn} \\
IPPP, Department of Physics, University of Durham, \\
Durham DH1 3LE, England, \\
E-mail: \email{christian.schwinn@durham.ac.uk}
}

 \abstract { We consider the resummation of soft and Coulomb gluons for
   pair-production processes of heavy coloured particles at hadron
   colliders, and discuss recent results on the construction of a
   basis in colour space that diagonalizes the soft function to all
   orders in perturbation theory and the determination of the two-loop
   soft anomalous dimension needed for NNLL resummations.  We 
   present results for the combined NLL resummation of soft
   gluon and Coulomb-gluon effects for squark-antisquark production at
   the LHC.
                
}
 \dedicated{
\hfill\\[1cm]
PITHA~09/21, 
IPPP/09/69,
DCPT/09/138,
SFB/CPP-09-84\hfill
September 18, 2009

}

\FullConference{The 2009 Europhysics Conference on High Energy Physics,\\
		 July 16 - 22 2009\\
		 Krakow, Poland}

\begin{document}

\section{Introduction}

Perturbative corrections to the partonic cross sections for pair
production of heavy coloured particles $H, H'$~(top quarks, squarks,
gluinos...) at hadron colliders contain terms of the form $[\alpha_s
  \ln^2\beta\,]^n$ (``threshold logarithms'') and $(\alpha_s/\beta)^n$
(``Coulomb singularity'') that are enhanced near the partonic
threshold $\hat s\approx 4 M^2$ but can be resummed to all orders in
perturbation theory.  (Here $\beta=(1-4M^2/\hat s)^{1/2}$, $M$ the
average heavy-particle mass and $\hat s$ the partonic centre of
mass-energy.)  Resummation of threshold logarithms at next-to leading
logarithmic accuracy~(NLL) was achieved in~\cite{Kidonakis:1997gm}
in the Mellin-moment formalism~\cite{Sterman:1986aj}. Coulomb
resummation has been performed for top, squark and gluino pair
production~\cite{Hagiwara:2008df,Kiyo:2008bv,Kulesza:2009kq}.  Here we
report on work~\cite{Beneke:2009rj} towards threshold resummation at
next-to-next-to leading logarithmic~(NNLL) accuracy and the
combination with Coulomb resummation.  We present results for a
combined momentum space~\cite{Becher:2007ty} resummation of NLL
threshold logarithms and leading Coulomb corrections for
squark-antisquark production at the LHC.

\section{Factorization, diagonalization and resummation}
Near threshold, the partonic cross sections of the subprocesses $p
p'\rightarrow HH'+X$ where $pp'\in\{qq, q\bar q, gg, gq,g\bar q\}$
satisfy the factorization formula~\cite{Beneke:2009rj} 
(with $E=\sqrt{\hat s}-2 M$)
\begin{equation}
\label{eq:fact}
  \hat\sigma_{pp'}(\hat s,\mu)
= \sum_{i,i'}H_{ii'}(M,\mu)
\;\int d \omega\;
\sum_{R_\alpha}\,J_{R_\alpha}(E-\tfrac{\omega}{2})\,
W^{R_\alpha}_{ii'}(\omega,\mu)
\end{equation}
 that separates hard, soft and potential effects and includes the
 leading Coulomb singularities $(\alpha_s/\beta)^n$ to all
 orders.\footnote{Eq.~\eqref{eq:fact} applies to S-wave production of
   the heavy particles. Note that (as mentioned
   in~\cite{Beneke:2009rj}) this formula does not include
   $\mathcal{O}(\alpha_s^2\log\beta)$ terms related to higher order
   potential effects and higher-dimensional soft functions that may
   account for the three parton correlations reported
   in~\cite{Ferroglia:2009ii}.  However, these NNLL effects do not
   affect the soft function $W_{ii'}^{R_\alpha}$ discussed here.} The
 sum extends over the irreducible colour representations in the
 decomposition $R\otimes R^\prime=\sum_{\alpha}R_\alpha$ of the
 final-state representations.  The potential function $J^{R_\alpha}$
 is proportional to the imaginary part of the non-relativistic
 zero-distance Coulomb Green function quoted
 e.g. in~\cite{Kiyo:2008bv}.  The soft function $W^{R_\alpha}_{ii'}$
 is defined in terms of soft Wilson lines (see eq.~(1.6)
 of~\cite{Beneke:2009rj}) and is related to the threshold logarithms
 while the $H_{ii'}$ encode the partonic hard-scattering processes.
 The indices $i,i'$ denote the elements $c^{(i)}_{\{a\}}$ of a basis
 of colour structures. As we have shown in~\cite{Beneke:2009rj}, the
 soft function is diagonal in the basis
\begin{equation}
\label{eq:prod-basis}
  c_{\{a\}}^{(i)}=\text{dim}(r_\alpha)^{-\frac{1}{2}}\;
  C^{r_\alpha}_{\alpha a_1a_2} C^{R_{\beta}\ast}_{\alpha a_3a_4},
\end{equation}
where the $C$ are Clebsch-Gordan coefficients combining the
initial~(final) state representations $r,r'$ ($R,R'$) to an
irreducible representation $r_\alpha$ ($R_\beta$) and the index $i$
labels pairs $P_i=(r_{\alpha}, R_{\beta})$ of equivalent initial and
final state representations.  For  quark-antiquark and
gluon initiated squark-antisquark production
the allowed pairs are $P_i\in
\{(1,1),\; (8,8) \}$ and $P_i\in
\{(1,1),\; (8_S,8),\;(8_A,8) \}$.

In the diagonal basis, the  soft function in position
space satisfies an evolution equation~\cite{Beneke:2009rj}
\begin{equation}
\label{eq:rge-soft}
\frac{d}{d\ln\mu} \hat W^{R_\alpha}_{ii}(z_0) = 2\left(
(\Gamma_{\text{cusp}}^{r}+\Gamma_{\text{cusp}}^{r'})\ln\left(\frac{i
  z_0\mu e^{\gamma_E}}{2}\right)
-(\gamma_{H,s}^{R_\alpha}+\gamma^r_s+\gamma_s^{r'})\right) \hat
W^{R_\alpha}_{ii}(z_0),
\end{equation}
similar to that of the soft function in the Drell-Yan
process~\cite{Korchemsky:1993uz}.  Threshold logarithms can be
resummed~\cite{Korchemsky:1993uz,Becher:2007ty} by evolving the soft
function using~\eqref{eq:rge-soft} from a scale $\mu_s$ characteristic
for the soft radiation to a scale $\mu_f$ where the parton
distribution functions are evaluated.  Using an analogous equation,
the hard function is evolved from a scale $\mu_h\sim 2M$ to $\mu_f$.
The anomalous dimensions $\Gamma_{\text{cusp}}^{r}$ and $\gamma^r_s$
for a massless particle in the SU(3) representation $r$ are known with
the accuracy required for NNLL resummations.  The soft anomalous
dimension of the heavy particle system in the representation
$R_\alpha$, $\gamma_{H,s}^{R_\alpha}$, has been given at one-loop for
some examples~\cite{Kidonakis:1997gm,Kulesza:2009kq}. The result up to
two-loops, as required for NNLL resummations, was extracted
in~\cite{Beneke:2009rj} from results of~\cite{Becher:2009kw} and
confirmed by~\cite{Czakon:2009zw} for top production (with
$C_{R_\alpha}$ the quadratic Casimir invariant for the representation
$R_\alpha$):
\begin{equation}
\label{eq:gamma-H}
\gamma_{H,s}^{R_\alpha}=\frac{\alpha_s}{4\pi}\left(-2 C_{R_\alpha}\right)
+\left(\frac{\alpha_s}{4\pi}\right)^2C_{R_\alpha}\left[- C_A
\left(\frac{98}{9}-\frac{2\pi^2}{3}+4\zeta_3\right)
+\frac{40}{18} n_f\right]+\mathcal{O}(\alpha_s^3).
\end{equation} 

\section{Combined soft and Coulomb resummation for squark-antisquark production}
  The factorization formula~\eqref{eq:fact} allows a combined
  resummation of soft- and Coulomb effects using the momentum space
  solution~\cite{Becher:2007ty} to the evolution
  equation~\eqref{eq:rge-soft} and the resummed Coulomb Green function
  $J$. 
  We give here an analytical
  result for the partonic cross section that includes soft corrections
  resummed to NLL accuracy (the first term in square brackets), single
  Coulomb exchange interfering with resummed NLL soft radiation (the
  second term in square brackets) and higher-order Coulomb corrections
  $(\alpha_s/\beta)^n$ resummed without soft radiation (the last
  term):
\begin{equation}
\label{eq:fact-resum-c1}
\begin{aligned}
\hat\sigma^{\text{NLL}\otimes C}_{pp'}=
\sum_{i,R_{\alpha}} \,\hat\sigma^{i,(0)}_{pp'}
\left\{ U^{R_\alpha}_i
\left(\frac{E e^{-\gamma_E}}{M}\right)^{2 \eta}
\left[\frac{\sqrt{\pi}}{2\Gamma(2\eta+\frac{3}{2})}\,
+\frac{\kappa_{R_\alpha}}{\Gamma(2\eta+1)} 
\right]+2\kappa_{R_\alpha}\text{Im}[\psi(1-i\kappa_{R_\alpha})]
\right\}.
\end{aligned}
\end{equation}
Here $\kappa_{R_\alpha}=(-C_{R_\alpha}/2)
\alpha_s(\mu_C)\sqrt{2m_{\text{red}}/E}$ with the reduced mass
$m_{red}$ of the heavy particle pair, $C_{1}=-C_F$, $C_8=1/(2N_C)$,
$\sigma^{i(0)}$ are the tree cross sections for the singlet and octet
colour channels~\cite{Langenfeld:2009eg} at threshold, $\eta=2
a_{\Gamma}(\mu_s,\mu_f)$ and ($\gamma_i^{V}$ and $\gamma^{\phi,r}$ are
given in~\cite{Beneke:2009rj}):
\begin{align}
  U_i^{R_\alpha}&=
\exp[4  S(\mu_h,\mu_s)
 -2a_i^{V}(\mu_h,\mu_s) +2 a^{\phi,r}(\mu_s,\mu_f)+2 a^{\phi,r'}(\mu_s,\mu_f)]
\left(\tfrac{4M^2}{\mu_h^2}\right)^{-2a_\Gamma(\mu_h,\mu_s)}\nonumber\\
S(\mu_h,\mu_s) &= -\int_{\alpha_s(\mu_h)}^{\alpha_s(\mu_s)}d \alpha_s 
\frac{\Gamma^r_{\text{\text{cusp}}}(\alpha_s)
  +\Gamma^{r'}_{\text{\text{cusp}}}(\alpha_s)}{2\beta(\alpha_s)}
\int_{\alpha_s(\mu_h)}^{\alpha_s}\frac{d \alpha_s^{'}}{\beta(\alpha_s^{'})} 
\, ,\nonumber\\
a_{\Gamma}(\mu_a,\mu_b) &= -\int_{\alpha_s(\mu_a)}^{\alpha_s(\mu_b)} d \alpha_s 
\frac{\Gamma^{r}_{\text{\text{cusp}}}(\alpha_s)
+\Gamma^{r'}_{\text{\text{cusp}}}(\alpha_s)}{2\beta(\alpha_s)}
\, , 
\qquad a^{V}_i (\mu_a,\mu_b) = -\int_{\alpha_s(\mu_a)}^{\alpha_s(\mu_b)} d \alpha_s 
\frac{\gamma_i^{V}(\alpha_s)}{\beta(\alpha_s)}.
\end{align}

In order to calculate the squark-antisquark production 
cross section at the LHC, we match to the
fixed-order NLO calculation~\cite{Beenakker:1996ch} in the
parameterization of \cite{Langenfeld:2009eg} (see
e.g.~\cite{Kulesza:2009kq}) and convolute the partonic cross
section~\eqref{eq:fact-resum-c1} with the MSTW08NNLO
parton-distribution functions.  We choose $\mu_s$ as the scale $\tilde
\mu_s$ that minimizes the fixed-order one-loop soft
corrections~\cite{Becher:2007ty}, resulting in $\tilde{\mu}_s =
123-455$ GeV for $m_{\tilde{q}}=200-2000$ GeV. In the quantity
$\kappa_{R_\alpha}$ we set $ \mu_C= 2M\,\text{max}\{\beta,
\alpha_s(\mu_C)\}$, motivated by the momentum transfer $|\vec k|\sim M
\beta\sim M\alpha_s$ involved in the Coulomb corrections and the form
of the known higher order corrections to the Coulomb Green function.
In the results below, we identify the hard- and factorization scales,
$\mu_h=\mu_f$.  The black~(solid) line in the left plot in
figure~\ref{fig:squarks} shows the corrections from soft and
Coulomb-gluon effects as described by~\eqref{eq:fact-resum-c1}
relative to the fixed-order NLO cross section. The blue (long-dashed)
line shows the NLL soft-gluon corrections alone (i.e. only the first
term in the square bracket in~\eqref{eq:fact-resum-c1}), in
qualitative agreement with results from NLL resummation in Mellin
space~\cite{Kulesza:2009kq}.  The red~(dot-dashed) curve includes the
effect of Coulomb-resummation added to the soft NLL corrections
(without soft-Coulomb interference). Our choice for $\mu_C$ results in
larger corrections than those found in~\cite{Kulesza:2009kq} for
$\mu_C=\mu_f$. Finally we compare to the
NNLO$_{\text{approx}}$-results in eqs.~(17-18)
of~\cite{Langenfeld:2009eg} which include soft-Coulomb interference
and the running coupling in the Coulomb potential at fixed order,
together with NNLL soft corrections and further two-loop Coulomb
corrections.  For the magenta~(dotted) curve the tree-level scaling
functions $f^{(00)}$ at threshold have been used (as $\sigma^{i(0)}$
in~\eqref{eq:fact-resum-c1}) whereas in the green~(dashed) curve the
full tree-level result is used as in~\cite{Langenfeld:2009eg}. The
relative corrections obtained from~\eqref{eq:fact-resum-c1} and the
NNLO$_{\text{approx}}^{\text{NR}}$ result include different higher
order effects and hence differ by $20-35\%$.  The right plot in
figure~\ref{fig:squarks} shows the reduced factorization scale
dependence of the NLL  compared to the NLO cross section. The red band
accounts for the variation of the soft scale between
$\tilde\mu_s/2<\mu_s<2\tilde \mu_s$. 
 \FIGURE{
  \includegraphics[width=0.475\textwidth]{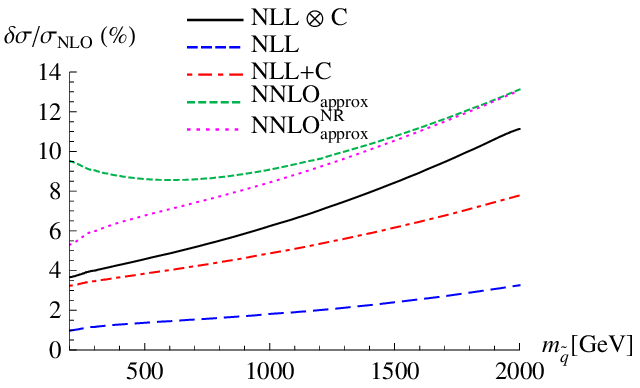}
  \includegraphics[width=0.45\textwidth]{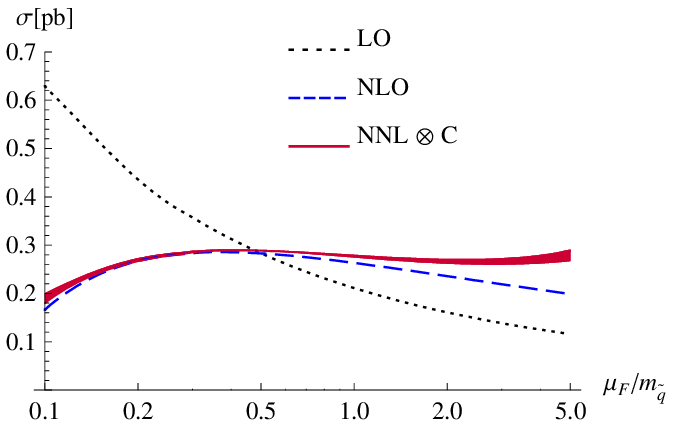}
    \caption{ Squark-antisquark production cross section for
      an $14$~TeV LHC and a gluino mass $m_{\tilde g}=1.25 m_{\tilde
        q}$. Left: Relative corrections in various
      approximations. Right: factorization scale dependence for
      $m_{\tilde q}=1$~TeV. 
    }
\label{fig:squarks}
}

\end{document}